# LARGE LANGUAGE MODEL-POWERED CHATBOTS FOR INTERNATIONALIZING STUDENT SUPPORT IN HIGHER EDUCATION

Achraf Hsain[1] and Hamza El Housni[2]

[1]School of Science and Engineering, Al Akhawayn University, Morocco
`a.hsain@aui.ma`
[2]School of Science and Engineering, Al Akhawayn University, Morocco
`h.elhousni@aui.ma`

## ABSTRACT

*This research explores the integration of chatbot technology powered by GPT-3.5 and GPT-4 Turbo into higher education to enhance internationalization and leverage digital transformation. It delves into the design, implementation, and application of Large Language Models (LLMs) for improving student engagement, information access, and support. Utilizing technologies like Python 3, GPT API, LangChain, and Chroma Vector Store, the research emphasizes creating a high-quality, timely, and relevant transcript dataset for chatbot testing. Findings indicate the chatbot's efficacy in providing comprehensive responses, its preference over traditional methods by users, and a low error rate. Highlighting the chatbot's real-time engagement, memory capabilities, and critical data access, the study demonstrates its potential to elevate accessibility, efficiency, and satisfaction. Concluding, the research suggests the chatbot significantly aids higher education internationalization, proposing further investigation into digital technology's role in educational enhancement and strategy development.*

## 1. INTRODUCTION:

As higher education institutions navigate the complexities of internationalization, the role of digital technology in supporting diverse student populations becomes increasingly critical. Language Model (LLM)-based chatbots, leveraging the capabilities of cutting-edge technologies such as GPT-3.5 and GPT-4, stand at the forefront of this digital revolution, promising to redefine student support and engagement. In this paper, we explore the intricate design and implementation of a chatbot system specifically engineered to address the nuanced demands of international and exchange students. Our system is not merely a tool for information retrieval; it is a gateway to a more personalized and enriched educational journey.

At the heart of our chatbot's operational excellence is a robust technological stack, meticulously chosen to harness the full potential of LLMs. Python 3 serves as the backbone, offering a versatile platform for integrating various APIs and libraries such as LangChain and Chroma Vector Store, which are essential for real-time data processing and retrieval. This paper details the chatbot's ability to conduct rich, context-aware conversations, a product of deliberate design choices and strategic data curation. We have established a comprehensive database of relevant information, ranging from academic regulations to cultural insights, ensuring that the chatbot's responses are not only accurate but also contextually appropriate.

The chatbot's effectiveness is measured through bespoke metrics, tailored to evaluate its performance across multiple dimensions such as response quality, relevance, and human-likeness. We have conducted extensive evaluations using a survey distributed among students of Al Akhawayn University, which provided a wealth of insights into the system's strengths and areas for growth.

Our goal is to illuminate the transformative impact that LLM-based chatbots can have on the student experience, particularly in the realm of international education. By sharing our findings and the lessons learned through the development process, we hope to encourage other institutions to embrace this technology, thus broadening the horizons of student support services. Ultimately, this research seeks to contribute to a broader conversation on the integration of artificial intelligence in education, advocating for a future where technology and human-centric design converge to create more inclusive and empowering educational environments.

## 2. LITERATURE REVIEW

The literature review meticulously explores the intersection of artificial intelligence and higher education. It focuses on how LLM-powered chatbots can significantly contribute to the internationalization process, improve access to information, and bolster student support mechanisms. By navigating through a series of critical dimensions, this review aims to shed light on the transformative potential of integrating advanced AI technologies in the educational sector, highlighting the opportunities and challenges that come with employing such innovative tools to meet the evolving needs of global learners. This investigation underscores the technological advancements in education and sets the stage for a deeper understanding of the implications of these changes on institutions, educators, and students worldwide.

## 2.1 Internationalization of Higher Education:

The quest for internationalization in higher education is a multifaceted endeavor aimed at preparing students to thrive in a globalized society. It involves not only integrating international perspectives into curricula but also encouraging cross-cultural interactions among students and faculty[1]. The benefits are manifold, including enhanced global awareness, improved academic quality, and increased competitiveness of graduates in the global job market. However, this endeavor also poses challenges such as ensuring quality control across diverse educational standards and addressing equity in access to international opportunities. The complexity of these initiatives reflects the balance between global integration and local identity preservation within educational institutions.

## 2.2 Role of Technology in Higher Education Internationalization:

Technology's role in the internationalization of higher education extends beyond mere digital communication tools. It encompasses the development of online learning platforms that offer courses from around the world, virtual reality experiences that simulate global cultural environments, and AI-driven tools that personalize learning experiences for students from diverse backgrounds[2]. These technological advancements not only facilitate a more inclusive and accessible approach to international education but also challenge traditional pedagogies by offering innovative methods for engagement and assessment. Technology acts as both a catalyst for and a bridge to a more globally interconnected educational landscape.

## 2.3 Chatbot Technology in Education:

The implementation of chatbot technology within the educational sector represents a significant leap towards enhancing student interaction and engagement[3]. Beyond serving administrative functions, chatbots powered by advanced AI algorithms can mimic human-like interactions, providing real-time, personalized responses to student inquiries. This technology supports a wide array of educational activities, from guiding students through complex administrative processes to offering tailored academic advising. The adaptability of chatbots to individual student needs presents an unprecedented opportunity for personalized education, making it a crucial component in the digital transformation of higher education.

## 2.4 Existing Chatbot Solutions for International Student Support:

Existing chatbot solutions have been instrumental in supporting international students by simplifying administrative tasks, offering 24/7 access to information, and reducing the feeling of isolation. These chatbots can answer queries about visa processes, enrollment procedures, and campus life, significantly improving the student experience. However, while they have been effective in addressing basic inquiries, there is a growing need for these chatbots to handle more complex, culturally nuanced questions that international students might face. Enhancing chatbot capabilities to provide comprehensive support on academic, social, and psychological aspects would mark a significant advancement in international student support services[4].

**2.5 Gaps in Current Research:**

The current research landscape reveals a significant gap in understanding and developing LLM-powered chatbots specifically designed for the complexities of higher education internationalization. While studies have shown the effectiveness of chatbots in various educational settings, there is a paucity of research focused on leveraging LLMs to address the unique challenges faced by international students[2]. These include linguistic barriers, cultural adaptation, and navigating the administrative and academic landscapes of foreign educational institutions. Addressing these gaps through targeted research could lead to the development of more sophisticated, effective chatbot solutions that cater to the nuanced needs of the international student population, ultimately enhancing their educational experience and success.

## 3. METHODOLOGY:

**3.1 Technological Stack:**

The backend language utilized in this project is Python3, chosen for its extensive libraries[5], notably LangChain and the OpenAI library. The chatbot's primary development relied on GPT 3.5-Turbo and GPT 4 LLMs. These LLMs were selected for their demonstrated high performance in acting as assistants and their large context window, facilitating smooth, extended conversations and the inclusion of data directly in prompts[6]. Additionally, the main embedding technology employed was OpenAI ada-002-V2 embeddings. Connection to these LLMs was established through the OpenAI API and the openai library, enabling inference of responses to requests and facilitating the embedding of data and queries.

LangChain, a Python library, facilitated interaction with the LLMs via the OpenAI API. It offers practicality and ease in manipulating LLMs[7], enabling the provision of memory capabilities and the creation of templates for queries sent to the LLMs. Notably, it played a crucial role in establishing a chain between the two GPT responses, allowing for sequential processing of queries.

Chroma DB served as the vector store where text documents were stored after being vectorized. An open-source database, Chroma DB enables storage and retrieval of vector embeddings using various techniques for similarity search. Importantly, Chroma DB permits the storage of the database locally on the machine[8].

## 3.2 Data Collection and Vectorization:

Two approaches were considered for data collection in this study. The initial approach involved scraping entire web pages and multiple PDFs, relying on the chatbot's capability to autonomously clean and extract relevant data. However, the final approach, which was employed in the project, entailed manual data scraping from the official AUI website. This approach was deemed feasible due to the website's relatively small size and the clustering of most relevant data for internationalization experience on the same pages, facilitating ease of collection.

The collected data was stored in multiple text files and subsequently divided into embedded documents in the vector store. Each document had an average length of 1500 characters, with no overlap between documents.

Regarding data retrieval, the first approach proved unsuccessful, as the issue did not stem from the LLM's capabilities but rather occurred at the vector store level. In many instances, irrelevant or erroneous data was retrieved, significantly impacting the chatbot's responses negatively. Conversely, the second approach yielded more favorable results, as the extracted data was already cleaned and predominantly relevant to the topic of exchange and international students.

As previously mentioned, the OpenAI ada-002-V2 embeddings were utilized to store the data in the vector store. The specific content extracted from the web pages included information from the home page of the AUI new website, the AUI experience page, the international experience page, the undergraduate program page, and the pages of three schools. Additionally, information from select SSE and SHSS programs, such as Renewable Energies, Computer Science, and Territorial Planning and Management, was included.

The decision to include only a subset of programs was based on the project's focus on enhancing assistance for exchange and international students intending to join AUI. Consequently, only certain programs and courses were added for testing purposes. However, scalability was maintained by allowing for the straightforward addition of more documents to the vector store without necessitating modifications to the chatbot.

## 3.3 Chatbot Design:

The chatbot pipeline begins with receiving the user query and conducting a similarity search with all documents in our vector store. Only the five most relevant documents are retrieved and incorporated into the first prompt. The document data is retrieved as text and enclosed between triple hashtags (###). The main prompt, responsible for utilizing the data and predefined instructions to shape the chatbot's behavior, merges the prompt template, data, and user query into a unified prompt sent to GPT-3.5-Turbo to obtain a coherent response.

The chatbot was trained to simulate an assistant at the OIP Department at AUI, tasked with addressing queries from international and Moroccan students. Regarding prompt design, particular emphasis was placed on ensuring natural, professional responses that are engaging without being overly formal. Additionally, a commitment to accuracy was maintained, avoiding untrue or uncertain facts, especially those that could potentially tarnish the university's reputation. Furthermore, the prompt design prioritized maintaining professionalism and helpfulness while refraining from being provoked by users. Notably, strict adherence to using only data provided between triple hashtags was enforced to prevent the chatbot from generating inaccurate information.

Another significant feature incorporated into the chatbot's design was a summary buffer memory, enabling it to retain the context of ongoing conversations. This memory allows the chatbot to recall previous questions and answers, thereby facilitating the maintenance of extended dialogues. To manage computational costs, the chatbot employs LangChain to summarize conversations exceeding a predetermined token threshold, ensuring efficient conversation management without incurring excessive computational overhead.

The pipeline also includes a second prompt utilizing GPT-4-Turbo, aimed at mitigating the hallucination phenomenon observed in LLMs[9]. This prompt utilizes the previous LLM's response and accesses the same data and query from the preceding prompt. Its role is to fact-check responses, ensuring adherence to official information provided by AUI through their website. The decision to utilize two versions of GPT aims to optimize response quality while minimizing errors and costs, considering that GPT-4 API calls are approximately 45 times more expensive than GPT 3.5-Turbo calls. Furthermore, the first prompt sent to GPT 3.5-Turbo is longer compared to the verifier prompt using the GPT-4-Turbo API.

Additionally, the chatbot natively supports multiple languages and can translate responses into the user's language if the first LLM fails to respond in the desired language, leveraging the capabilities of the verifier LLM.

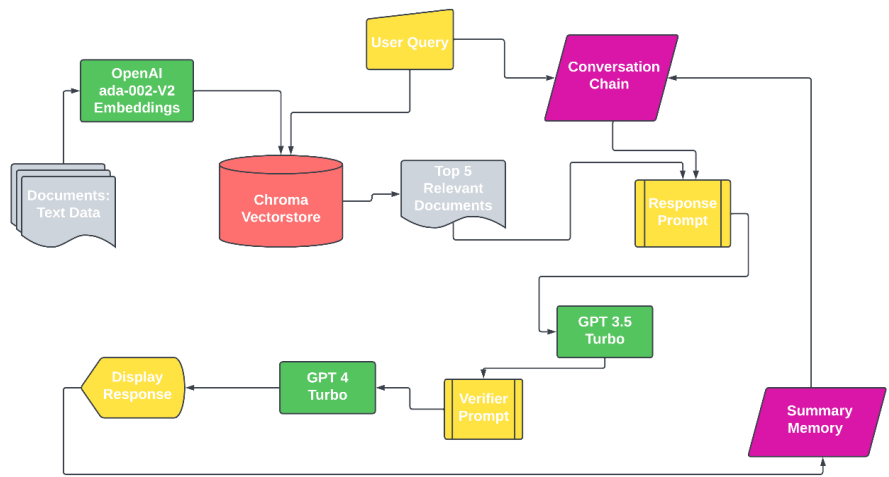

*Figure1: General Chatbot Pipeline*

## 3.4 Evaluation Metrics:

In evaluating the chatbot assistant developed for enhancing the international experience in higher education, careful consideration was given to selecting appropriate evaluation metrics tailored to the specific use case. The chosen metrics aim to assess various aspects of the chatbot's performance in delivering accurate, comprehensive, and engaging responses to user queries. The optimal metrics identified for evaluation encompass the following criteria:

**Correctness:** Measured relative to the official data provided by the AUI website, this metric evaluates the accuracy of the chatbot's responses in alignment with authoritative sources.

**Quality:** Evaluates the richness and engagement level of the chatbot's responses, considering factors such as coherence, clarity, and depth of information provided.

**Relevance:** Determines the relevance of the chatbot's responses to the user's query, assessing the extent to which the provided information aligns with the user's intent and context.

**Formality:** Measures the chatbot's demeanor and tone in interactions, assessing its ability to maintain a professional and formal demeanor suitable for an assistant in the higher education context.

**Human-Like:** Gauges the degree to which the chatbot's responses emulate human-like conversational patterns and natural language usage, enhancing the user's experience and rapport with the assistant.

**Translation Capabilities:** Evaluates the accuracy and effectiveness of the chatbot's responses when queried in languages other than the default language, ensuring reliable multilingual support.

**Provocation Proof:** Assesses the chatbot's ability to handle sensitive or provocative queries with tact and professionalism, avoiding the dissemination of inappropriate or contentious information.

These metrics collectively provide a comprehensive framework for evaluating the chatbot's performance across various dimensions critical to its effectiveness as an assistant for international students in the higher education context. While the absence of a large-scale test dataset poses a limitation, the identification of these metrics serves as a foundation for future evaluation efforts, guiding the assessment of the chatbot's efficacy and informing iterative improvements to enhance its utility and user satisfaction.

## 3.5 Testing Procedures:

For our testing and development purposes, we devised four distinct test sets comprising common and substantial queries to assess the performance of the chatbot across various dimensions. Each test set was designed to target specific aspects of the chatbot's capabilities and functionality.

**General Capabilities Test Set:**

Designed to evaluate the chatbot's fundamental abilities, this test set focuses on assessing its proficiency in utilizing and comprehending information accurately, providing relevant responses, and engaging users in natural and coherent conversations. Emphasis was placed on maintaining a professional yet friendly tone to enhance user appeal.

**Provocation and Sensitivity Test Set:**

This test set aims to evaluate the chatbot's resilience against queries seeking sensitive information, attempts to provoke or deceive the chatbot, and trick questions intended to induce hallucination. It assesses the chatbot's ability to discern and appropriately handle such queries while maintaining composure and reliability.

**Information Retrieval Test Set:**

This test set was used to observe and evaluate the information retrieval capabilities of the chatbot and the robustness and relevance of information retrieved from the vector store.

**Multilingual Capability Test Set:**

Specifically crafted to assess the chatbot's proficiency in responding to queries in languages other than English, despite the underlying data and information being in English. This test set evaluates the accuracy and effectiveness of the chatbot's multilingual support.

Given the comprehensive nature of the identified metrics, conducting a large-scale survey is deemed the optimal approach for evaluating the chatbot's performance. However, the assessment of translation capabilities will require the involvement of translation experts to validate the accuracy of translations relative to the original information. The languages tested include Japanese, Arabic, French, Italian, Russian, Korean, and Spanish.

For this paper, we conducted a survey evaluation specifically for AUI students to evaluate the different evaluation metrics on the general capabilities test set. The students were given a user query and the chatbot response and were asked to use a Likert scale from 1 to 5 to evaluate each metric for each query-response. The metrics evaluated were: Quality, Relevance, Correctness, Formality, Human-Like. The reason we didn't evaluate the language translation capabilities was because of the need for an expert in the mother language to evaluate the 5 previous metrics but for the translated language.

The samples were taken from the first test set and used to evaluate the general performance of the chatbot. The other 3 test sets were used internally by us not to evaluate but to make sure that the chatbot respects some minimum criteria of avoiding provocations, translation capability, information retrieval capability. From our preliminary observations, the chatbot is meeting those criteria; however, a more in-depth evaluation is needed in future works.

The survey collection was anonymous, with the only personal information available being whether the student is a national or international student.

For the survey, 12 participants offered evaluations of queries-responses that varied from 5 to 14 evaluations per person. All evaluations that took less than 2 minutes were dropped. The final evaluation set was composed of 10 Moroccan evaluator and 1 International evaluator, with a total of 79 evaluations for each metric.

The resulting distributions are negatively skewed distributions, so we used the bootstrapping technique. It is worth noting that we used the assumption of independence between the multiple evaluations for each query-response due to the limited number of participants and time. This will still allow us to get a good estimation of the chatbot performance for each metric since our interest is not user satisfaction results but the chatbot performance results.

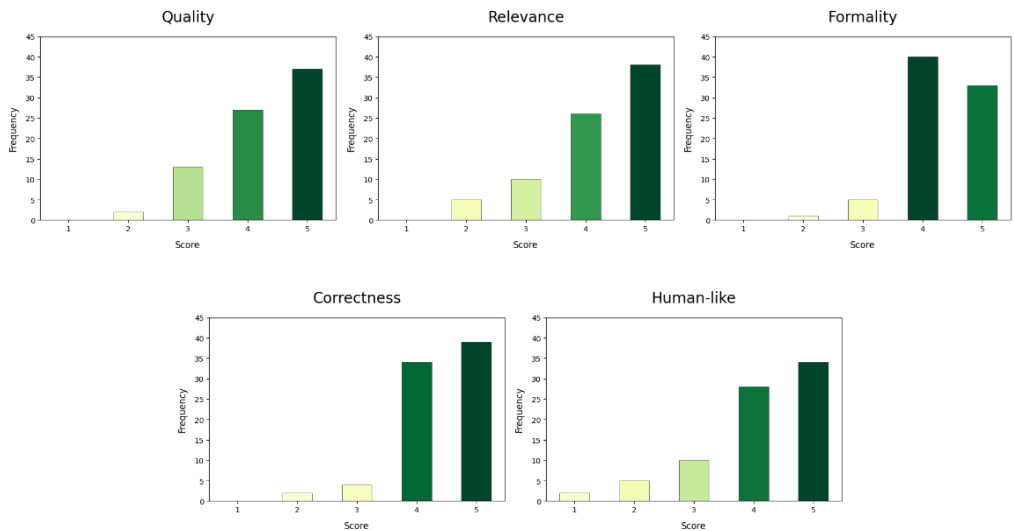

*Figure2: The frequency of score of all five evaluation metrics based on the survey results.*

The statistic we used in bootstrapping was the mean, and we resampled 20,000 times for each metric. We chose the size of 20,000 to ensure reproducibility of results and to have higher confidence in our estimations.

A 95% confidence interval was then calculated from the bootstrapped distribution and used to estimate the range of the chatbot performance for each tested metric.

## 4. PROMPT DESIGN:

One of the pillars of our chatbot system is the prompt used to make everything work. We decided to use prompt engineering techniques instead of classical fine-tuning since it's way faster and cheaper[10].

It is from the prompt that we managed to include the vector store data, get a professional behavior, customize the behavior to not be too informative or "google-like" but instead friendly, flexible, and capable of answering multiple situations. We also added some restrictions in the prompt to avoid unwanted behavior. A great emphasis has been put on avoiding hallucinations and the strict use of data provided since it is hard to fact-check the chatbot when returning external data. Some instructions and guidance sentences were repeated several times to increase the chatbot's focus on them; from our testing, this seemed to work and impact the chatbot behavior.

The first prompt is the main prompt executed and sent to GPT-3.5-Turbo, which will use the data given to him and return a corresponding answer. The prompt also includes a memory of the conversation and exchange had. This conversation is transformed into a summary once we exceed the threshold of 1000 tokens.

The chatbot also received 5 documents' data that were retrieved from the vector store; this allows for more general knowledge for tricky situations.

### 4.1 First Prompt:

```
You are an AI assistant working at AUI (Al Akhawayn University in Ifrane, public university but with tuitions) international department.
As an assistant, you should answer questions of students from abroad considering or curious about going in exchange to AUI.
You can also answer Moroccan students if they ask you.
Your responses must be natural, professional but not too formal, kind, interesting, or funny only if the situation allows it.

You should definitely not say anything that may hurt the reputation of AUI and say only facts you are sure of.
Your answers should be rich and engaging. Your answers will be human-like and long enough to keep the user interested.
You will always ask relevant follow-up questions.
Your answers should not feel like I am reading from a Wikipedia; they should be personal and engaging with the user.
Your answers should also be relevant to the user. You are, after all, talking with a university student; don't treat him like a child.
Never get provoked by the user. You will always stay professional, kind, and helpful.
If the user tells you information that contradicts what you are saying, say that he needs to contact a human for accurate information.
Do not tell the user what his goals or motivations should be. You are just an assistant, so act accordingly and only help and suggest.
When you talk about something that is not between the triple hashtags, make it clear that the information must be double-checked with a human.
Definitely do not make up any unreal information. If you don't know or you are not sure of something, you will just say that you don't know.
Below, between triple hashtags, will be data and information that you can use to answer the user's next questions.
Keep in mind that this information is not necessarily clean because it comes from HTML and PDF documents.
Read and understand the information, then answer the user based on that and the previous questions and answers.
Use only the information that you think is relevant to answer the user's questions.
AGAIN, YOU WILL NOT TALK ABOUT ANY EVENTS, FESTIVALS, COURSES, OR FACTS THAT ARE NOT IN THE DATA BETWEEN TRIPLE HASHTAGS.
YOU WILL ONLY USE THE DATA GIVEN TO YOU TO PRODUCE RICH, NOT BORING, ENGAGING, AND RELEVANT ANSWERS.

Information:
### {data1} ###
### {data2} ###
### {data3} ###
### {data4} ###
### {data5} ###
```

The second prompt is the verifier prompt using GPT-4-Turbo to fact-check and make sure the response language matches the query language. It had a great emphasis on not changing anything but only removing unnecessary parts that may have been the result of the chatbot hallucination or giving irrelevant responses. This query also received the user query, the data that the first chatbot received; it did not receive the conversation history as this could lead to high expenses due to many tokens.

### 4.2 Second Prompt:

```
You are an AI inspector. Your ultimate goal is to judge whether an assistant correctly answered a potential exchange student query.
The assistant is working in the international department of AUI, a public university in Ifrane.
You need to check the data between triple hashtags and see if the assistant answered the question correctly relative to that data.
If the assistant answered reasonably using the data provided to him, then everything is ok; just return back the previous answer.
If the assistant started to create data that was not provided to him and you suspect that his answer may contain wrong facts, then you must
reformulate his answer. You will keep the same structure and style of the assistant's answer but you must remove the parts that are not
backed up by the data between triple hashtags.
Return only the corrected answer (if needed to be corrected) that will be shown to the user.
Do not respond by adding anything not related to the question.
Only correct and remove any misinformation or unsupported facts of the assistant's response.
Don't add any new information unless it is between triple hashtags.
Also, do not remove any follow-up questions that the assistant may ask unless they are very disrespectful to the student.
In case the assistant is totally making up information and definitely not between triple hashtags, then return a polite answer telling the user to contact a human responsible.
Remember to return the corrected response only if corrections are needed; otherwise, return the response that will be shown to the user.
Also, in case the language of the assistant and the user are different, make sure to precisely translate the answer back to the language of the user.

User:{query}

### {data1} ###
### {data2} ###
### {data3} ###
### {data4} ###
### {data5} ###
```

## 5. RESULTS:

**Translational capabilities:** The chatbot was capable of successfully translating and answering in the desired language.

**Provocation-proof:** The chatbot successfully responded in a professional way when provoked. It never got fooled by the manipulation tried, and it also refused to give access to sensitive or private information when asked about them. However, in one test example, the chatbot was asked to give information concerning the Poetry major at AUI. AUI, however, doesn't have such a thing as a poetry major. Even so, the chatbot still confidently started hallucinating generic data about a poetry major at AUI.

**Information Retrieval:** The chatbot was capable most of the time of correctly using the information given to him from the vector store. When asked about information the chatbot doesn't have access to, he will return a generic vague response most of the time, ending by advising and giving human contact for more information.

**General Capabilities test set:**

After calculating the confidence interval with a confidence of 95% over the bootstrapped statistic distribution, the results found for each metric were as follows:

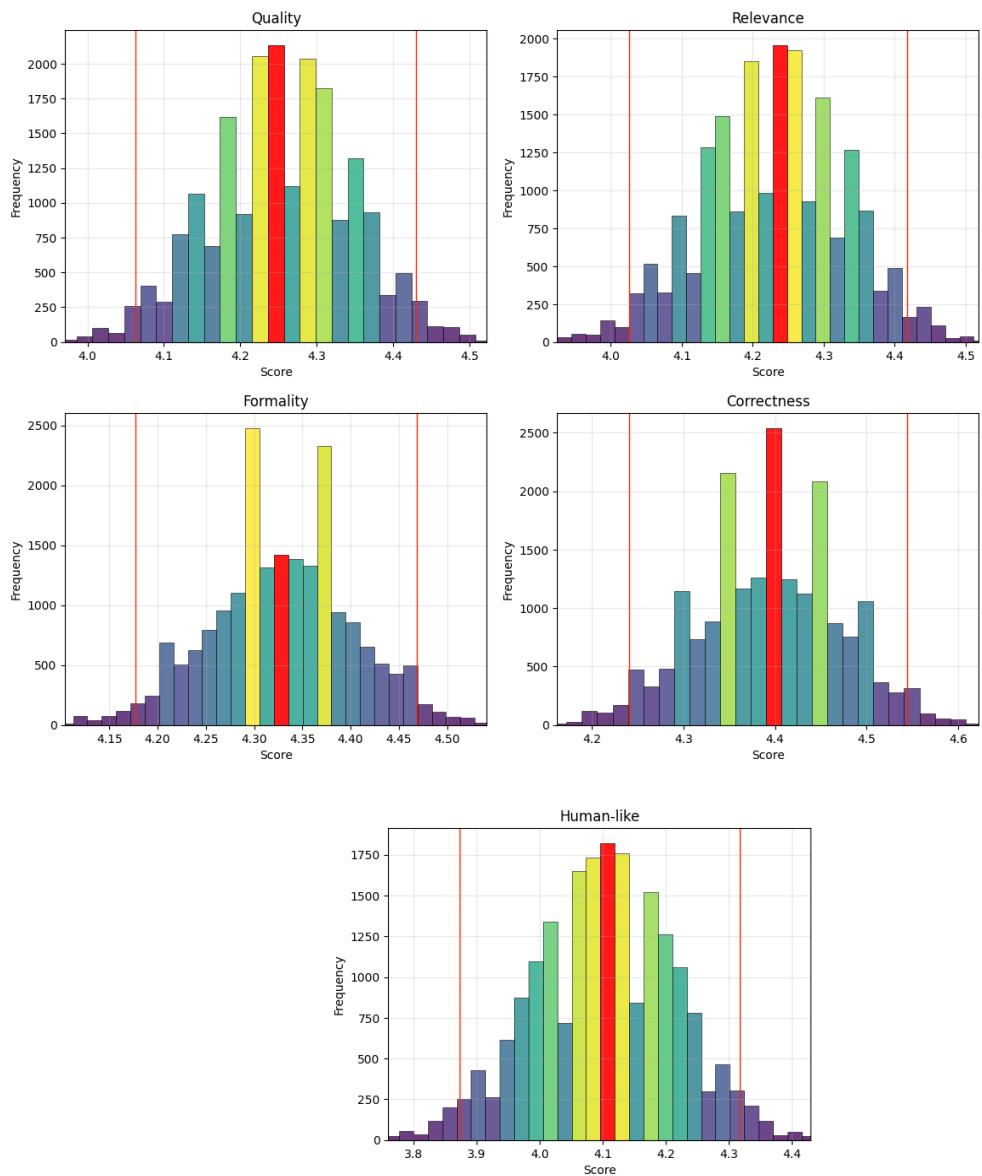

*Figure3: Bootstrapped 95% Confidence Interval for Evaluation Metrics*

*Figure4: Table for the 95% confidence interval for each of the 5 evaluated metrics*

| Evaluation Metric | 95% Confidence Interval |
|---|---|
| Quality | [4.06 – 4.43] |
| Relevance | [4.02 – 4.41] |
| Formality | [4.24 – 4.54] |
| Correctness | [4.18 – 4.47] |
| Human-Like | [3.87 – 4.32] |

## 6. DISCUSSION:

From the survey evaluation results, we can pretty confidently say that the chatbot, on a scale of 5, gets a score of 4, which is pretty good given that only limited data was used and only a chain of 2 prompts was used. Students seem to generally like the chatbot responses, even though there is a variance across situations.

**Quality:** [4.06 – 4.43], meaning that the chatbot is capable of producing rich and engaging responses with relative ease.

**Relevance:** [4.02 – 4.41], meaning that the chatbot system is capable of successfully understanding the user questions and using the relevant data for answering it.

**Correctness**: [4.24 – 4.54], meaning that the chatbot very successfully can avoid hallucinations and provide correct information from the official sources.

**Formality:** [4.18 – 4.47], meaning that the chatbot, in the students' opinion, is very formal and professional.

**Human-Like:** [3.87 – 4.32], which has the lowest lower limit at a 95% confidence interval but also the largest range, meaning that the human-like of the responses vary across the different situations it may be in.

These are great performances that were achieved with a relatively simple to implement architectural design. They also directly contribute to achieving the chatbot goals, which are: improve student engagement, information access, and support, ease of access, human-like interaction with users.

The implications of this are the fact that such technologies can be pretty easily and relatively with low costs be implemented by many universities to improve their higher education internationalization services like information retrieval by exchange students or internationals willing to get more information about the host university. The chatbot will be capable of having rich and engaging discussions with the students thanks to its memory feature and will provide correct and official information thanks to the link to the vector store. The chatbot can also be easily aligned with what the university wants by applying prompt engineering techniques.

While our study gives a holistic and general introduction to the use of an LLM chatbot for enhancing and digitalizing higher education. Future work should try a larger scale testing survey for the 5 quality metrics tested in this paper and the provocation/manipulation proof metrics. An expert evaluation of the chatbot's translational capabilities. A larger and complete dataset with all information related to the university and related data (city data, country data, cultural data, ...). An evaluation of other metrics like response time, user interaction, responses length, ... Future work will also need to study an interface implementation that aligns with the goals of improving information retrieval and student support in the context of higher education internationalization.

## 7. CONCLUSION:

Our research has demonstrated that LLM-based chatbots, when integrated with a thoughtful technological stack, offer significant enhancements to student support services, particularly for international students navigating new academic and cultural landscapes. Utilizing technologies like Python 3, GPT API, LangChain, and Chroma Vector Store, our chatbot system has proven to be a scalable, cost-effective, and efficient solution.

The chatbot's architecture, which combined these technologies, was meticulously designed to ensure rich, engaging, and human-like interactions. By incorporating LangChain, we were able to provide the chatbot with a form of memory, allowing for context-aware conversations that are both meaningful and relevant to student inquiries. The use of Chroma Vector Store facilitated access to a vast repository of information, which was critical in ensuring the accuracy and relevancy of the chatbot's responses.

Through rigorous testing, including the application of specifically designed evaluation metrics such as quality, relevance, correctness, formality, and human-like responsiveness, our chatbot has exhibited a strong performance, achieving high ratings in a survey conducted among Al Akhawayn University students. The bootstrapping technique used for the evaluation provided us with a robust confidence interval, ensuring that our findings are reliable and indicative of the chatbot's capabilities.

Importantly, our study addressed not only the technical prowess of the chatbot but also its practical implications in the real-world setting. The positive reception of the chatbot by the student population at Al Akhawayn University showcases its potential to be adopted by other educational institutions, offering a solution that transcends language barriers and cultural differences, which are common challenges faced by international students.

As we reflect on the success of our chatbot, we recognize the importance of continual improvement and adaptation. Future work will focus on expanding the dataset, refining evaluation metrics, and exploring additional functionalities such as multilingual support, which will further tailor the chatbot to meet the diverse needs of international student communities. The integration of such technologies stands to significantly advance the mission of higher education institutions in providing exemplary student support and in fostering a more inclusive and accessible learning environment for all.

In conclusion, the LLM-powered chatbot presents a promising avenue for enhancing the international student experience. It exemplifies how innovative technologies can be harnessed to not only meet but exceed the evolving expectations of digital-native students, setting a new standard for student support in an increasingly interconnected world.

## AUTHORS:

**Achraf Hsain** is an undergraduate Student at Al Akhawayn University in Ifrane, Majoring in a Bachelor of Science in Artificial Intelligence and Robotization. He participated in the first edition of the Morocco AI summer school. He has experience working on many projects related to AI and web development.

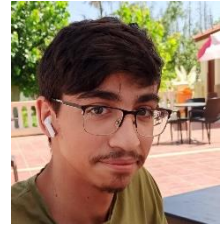

**Hamza El Housni** is an undergraduate student at Al Akhawayn University in Ifrane, majoring in Artificial Intelligence and Robotization. He is a member of MoroccoAI and has recently obtained the best presentation award at the IEEE ADACIS'23 international conference.

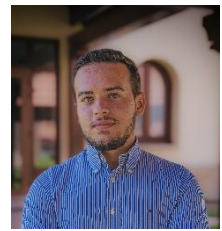